\documentclass[aps, prl, twocolumn, notitlepage, nofootinbib]{revtex4-1}

\usepackage{graphicx,amsmath,pstricks,subfig,float}

\DeclareMathOperator{\sgn}{sgn}

\begin{document}

\title{Life-hostile conditions in the early universe can increase the present-day odds of observing extragalactic life}

\author{S. Jay Olson}
 \email{stephanolson@boisestate.edu}
 \affiliation{Department of Physics, Boise State University, Boise, Idaho 83725, USA}

\date{\today}

\begin{abstract}
High-energy astrophysical events that cause galaxy-scale extinctions have been proposed as a way to explain or mollify the Fermi Paradox, by making the universe at earlier times more dangerous for evolving life, and reducing its present-day prevalence.  Here, we present an anthropic argument that a more dangerous early universe can have the opposite effect, actually increasing estimates for the amount of visible extragalactic life at the present cosmic time.  This occurs when civilizations are assumed to expand and displace possible origination sites for the evolution of life, and estimates are made by assuming that humanity has appeared at a typical time.  The effect is not seen if advanced life is assumed to always remain stationary, with no displacement of habitable worlds.
\end{abstract}

\maketitle

It has been proposed that the universe was a much more dangerous place for the evolution of intelligent life at earlier times~\cite{annis1999,vukotic2007,cirkovic2008a,vukotic2008,piran2014,li2015}.  The idea is attractive for understanding the Fermi Paradox, as it makes the Great Silence~\cite{brin1983} seem less paradoxical.  When applied to the universe as a whole (e.g. for the development of extragalactic SETI~\cite{annis1999b,zackrisson2015,wright2014b,griffith2015}), the idea comes at the cost of the mediocrity principle, as it puts us at a special cosmic time -- near the beginning of the time-of-arrival distribution for comparable life, $P_{TOA}(t)$.

This letter is to point out a context in which invoking a more dangerous early universe has the opposite effect, actually increasing estimates for the number of visible expansionistic Kardashev type iii+ civilizations~\cite{kardashev1964} (we denote expanding galaxy engineers as K3+), and improving the prospects for extragalactic SETI.  This can happen when two elements are simultaneously present in the reasoning used to make estimates:

\begin{enumerate}
	\item Advanced life is assumed to expand and eventually displace a large fraction of habitable worlds that would otherwise be potential sites for the appearance and evolution of life.  I.e. the rate at which expansionistic life appears at early cosmic times will affect the appearance rate of life at later cosmic times.
	\item A ``typicality assumption" for humanity's cosmic time of arrival is used to estimate the probability for advanced life to successfully evolve on a habitable world.  This is a form of anthropic reasoning that implicitly uses the self-sampling assumption~\cite{bostrom2002}.  It will not so much matter \emph{which} typicality assumption is used, so long as one is used.
\end{enumerate}

Because of the first element, the shape of the time-of-arrival distribution $P_{TOA}(t)$ (from which humanity is assumed to be randomly drawn) will respond in an amplified way to the deletion of early expansionistic civilizations.  The typicality assumption will then imply a larger, compensating estimate for the probability of advanced life to appear on habitable worlds that have not experienced a major extinction event, with the net effect that we estimate more K3+'s in our past light cone.

The described effect is not a pathological case -- indeed, it can be seen in previously published calculations that considered a number of increasingly severe evolutionary risk functions in the early universe, though the effect we describe here was not identified at the time~\cite{olson2017a}.  Although we will see this effect to be generic, it is not impossible to avoid.  It remains possible to back-engineer a risk function, such that we would estimate fewer visible K3+'s as a result of increased danger.

To be specific, let us express the appearance rate for advanced technological life (per unit comoving volume, per unit cosmic time) as $f(t) = \alpha \, F(t)$, where $ F(t)$ represents the rate at which approximately suitable planetary conditions have occurred (e.g. an Earthlike planet has appeared in the habitable zone of a sun-like star for sufficient time without experiencing a catastrophic extinction event), and $\alpha$ represents the conditional probability that advanced technological life will actually occur, given such planetary conditions.  The time variable here is cosmic time, and these rates are averaged across the universe.  We can then express the normalized time-of-arrival distribution as:
\begin{eqnarray}
P_{TOA}(t) = N^{-1} f(t)
\end{eqnarray} 
with the normalizing factor given by $N = \int_{0}^{\infty} f(t) dt$.  Now, if advanced life does not displace potential appearance sites, one can see that $P_{TOA}(t)$ does not depend on $\alpha$ -- it is canceled by the normalizing factor.  Invoking a more dangerous early universe changes the rate of suitable planetary conditions $F(t)$ such that early parts of $P_{TOA}(t)$ are suppressed, and the present cosmic time, $t_0$, becomes an unusually early time for us to have arrived.  This is a trade-off -- less observable life now (regardless of $\alpha$), in exchange for straining the mediocrity principle.  

However, if we assume that advanced life displaces a significant fraction of possible origination sites, the picture changes.  Let $g(t)$ be the fraction of possible origination sites that have \emph{not} been displaced.  Now, the time-of-arrival distribution is:
\begin{eqnarray}
P_{TOA}(t) = N^{-1} f(t) g(t)
\end{eqnarray} 
with the normalizing factor now given by $N = \int_{0}^{\infty} f(t) g(t) \, dt$.  This introduces an $\alpha$-dependence to $P_{TOA}$, since $g(t)$ should depend on the absolute appearance rate of advanced life.  

Before going any further, notice that models for $F(t)$ and $g(t)$ can be constructed and defended, based on specific assumptions about planet formation rates, technological limits and selection effects driving the behavior of expansionistic life.  Basic features of $F(t)$ have been inferred with reasonable confidence~\cite{lineweaver2001,loeb2016}, and the development of technology on the Earth suggests conservative bounds on what should be possible for a technologically mature species~\cite{merali2016}.  Drivers for expansionist behavior are discussed e.g. in~\cite{hanson1998,bostrom2014}.  Such issues are by no means settled, but one can imagine constraints and guiding principles that would partition the space of possibilities into a small set of consistent pictures.  By contrast, we are hopelessly uncertain over many orders of magnitude, where $\alpha$ is concerned~\cite{lacki2016}, with essentially no hope for a microscopic theory to inform us in the foreseeable future.  Because $P_{TOA}(t)$ depends on such an uncertain parameter, the mediocrity principle is one of the few tools at our disposal for making quantitative predictions.  To get estimates, we can demand that humanity has appeared at some ``typical" time, e.g. at the mean or median time of arrival, or $n$ standard deviations from the mean (for setting bounds).  This will fix $\alpha$ to a particular value, depending on the models used for $F(t)$ and $g(t)$.  One can then calculate the expected number of K3+'s that appear in our past light cone, for the set of assumptions.

Our argument turns on the observation that, in models with strong displacement, $F(t)$ by itself says almost nothing about our relative time of arrival. This is because \emph{any} plausible model for $F(t)$ can be paired with an equally-acceptable value of $\alpha$ (according to an uninformative prior, such as Lacki's Log-Log prior~\cite{lacki2016}) such that our appearance here at $t_0$ meets a typicality assumption.  If the typicality assumption is our only basis for selecting $\alpha$ (anthropic reasoning), then the effect of varying $F(t)$ by amount $\delta F(t)$ to make the early universe more hostile, e.g. to account for high-energy events, must be calculated \emph{while holding the typicality assumption constant}, meaning that $\alpha$ must be varied to compensate.

As a thought experiment to see how this reasoning will play out, imagine a proposal that the universe were extremely dangerous for a brief time, one \underline{m}illion years ago, causing a temporary, universe-wide halt to the appearance of K3+ life.  What is the net effect on estimates of the number of visible K3+'s from the Earth?  One million years is not enough time for light to arrive from distant, expansionistic galaxy engineers, so effects during the dangerous time itself cannot directly affect our estimates.  Nevertheless, $P_{TOA}(t)$ will be altered, because rapidly expanding civilizations would otherwise be displacing an increasing number of planets at all times in the future.  So this particular proposal means we will have to increase our estimate for $\alpha$ slightly to retain our chosen typicality assumption.  This increase in $\alpha$ will have observable consequences, because it affects the absolute appearance rate within our entire past light cone.  So the net effect of postulating danger in the past has been that we expect to see more advanced life, not less.

To examine this reasoning quantitatively, we functionally vary $F(t)$ by a small amount $ \delta F(t)$ for times before $t_0$, with $ \delta F(t)$ taking negative values.  To maintain the typicality assumption, this requires that we vary $\alpha$ by an amount $\delta \alpha$ to compensate.  We will examine two typicality assumptions, requiring that humanity (at $t_0$) has appeared at the mean or median time of arrival, so that we have $t_0 = N^{-1} \int_{0}^{\infty} t f(t) g(t) dt$ or $\frac{1}{2} = N^{-1} \int_{0}^{t_0} f(t) g(t) dt $ respectively, and then induce $\delta F(t)$, to make the early universe more hostile.  To first order, this implies the corresponding $\delta \alpha$ satisfies:
\begin{widetext}
\begin{eqnarray}
-\delta \alpha \, \int_{0}^{\infty} (t-t_0) \, F (t) \, \frac{\partial{g(t)}}{\partial{\alpha}} \, dt  =  \int_{0}^{\infty} dt' \, \delta F(t') \left[(t'-t_0)g(t') + \int_{0}^{\infty} (t-t_0) \, F(t) \, \frac{\delta g(t)}{\delta F(t')} dt       \right]
\end{eqnarray}
for the mean time, or
\begin{eqnarray}
-\delta \alpha \, \int_{0}^{\infty}  F (t) \, \frac{\partial{g(t)}}{\partial{\alpha}} \, \sgn(t_0 - t) \, dt  = \int_{0}^{\infty} dt' \, \delta F(t') \left[g(t') \sgn(t_0 - t') + \int_{0}^{\infty} \, F(t) \, \frac{\delta g(t)}{\delta F(t')} \, \sgn(t_0-t) \, dt   \right] 
\end{eqnarray}
for the median.
\end{widetext}

To go further, we need to be specific about a model for $g(t)$.  If $g(t)$ is dominated by the actions of species that use self-reproducing spacecraft to rapidly expand their domain in all directions at an intergalactic scale with speed $v$ (a generic kind of behavior for a sufficiently powerful agent wishing to maximize access to resources~\cite{bostrom2014}, and one that seems quite possible for mature technology~\cite{armstrong2013}), a model for the domain volume occupied by a species that appears at $t'$ is given by $V(t',t) = \frac{4 \pi}{3} \left( \int_{t'}^{t}  \frac{v}{a(t'')} dt'' \right)^3$ (with $a(t)$ the cosmic scale factor).  If such species appear at random in the universe at rate $f(t)$, and displace a large fraction of habitable worlds within their occupied volume, we obtain the following model for $g(t)$:
\begin{eqnarray}
g(t) = e^{- \int_{0}^{t} f(t') V(t',t) dt'}.
\end{eqnarray}
The collective time-dependence of such aggressively expanding civilizations resembles a cosmological phase transition~\cite{olson2014}, with $g(t)$ taking the form of the Guth-Tye-Weinberg formula~\cite{guth1980,guth1981}.  For such a model, the derivatives appearing in equations 3 and 4 are:
\begin{eqnarray}
\frac{\partial{g(t)}}{\partial{\alpha}} = \frac{g(t) \ln(g(t))}{\alpha} \\
\frac{\delta g(t)}{\delta F(t')} = - \alpha V(t',t) g(t) \Theta(t - t')
\end{eqnarray}
with $\Theta$ the Heaviside step function.

To determine the sign of $\delta \alpha$, one can check the sign of the integrals appearing in equations 3 and 4.  A sufficient condition for $\delta \alpha$ to be positive, is for $\delta F(t)$ to be negative and concentrated on times before $t_0$, i.e. ``making the early universe more dangerous."  The meaning is that deleting some of the early expanders with $\delta F(t)$ has the effect of pushing the average time of arrival to later cosmic times, while increasing $\alpha$ (increasing the number of expanding civilizations at all times) has an opposite, compensating effect, tending to saturate the universe more quickly.  The result is that $\delta \alpha$ must be positive to maintain either typicality assumption.

The expected number (average) of such expanding civilizations appearing within our past light cone (but not close enough to have reached our galaxy at speed $v$) is given by:
\begin{eqnarray}
E_{obs} = \left( \frac{c^3}{v^3} - 1 \right) \int_{0}^{t_0}  V(t', t_0)  f(t')   \, dt'
\end{eqnarray}
The \emph{change} in the expected number of such observable civilizations, as a result of making the early universe more hostile, is given by:
\begin{equation}
\delta E_{obs} = \left( \frac{c^3}{v^3} - 1 \right) \int_{0}^{t_0}  V(t', t_0)  \left[ \delta \alpha \, F(t') + \alpha \, \delta F(t') \right]   \, dt'.
\raisetag{\baselineskip}
\end{equation}
The negative $\delta F(t')$ term decreases the visible number, while $\delta \alpha$ increases it.  To see the effect of varying $F$ impulsively at a particular time $\bar{t}$, take $\delta F(t) = - \epsilon \, \delta(t - \bar{t})$, where $\epsilon$ is positive and $\bar{t}$ is earlier than $t_0$.  Then we express the change in the number of visible expanding civilizations as a function of $\bar{t}$:
\begin{eqnarray}
\delta E_{obs}(\bar{t}) = \epsilon \, E_{obs} Y(\bar{t}) - \epsilon \, \alpha  \left( \frac{c^3}{v^3} - 1 \right) V( \bar{t},t_0)
\end{eqnarray}  
with $Y(\bar{t})$ given by
\begin{equation}
Y(\bar{t})_{\mathrm{mean}} = \frac{- (\bar{t} - t_0)g(\bar{t})  + \alpha \int_{\bar{t}}^{\infty} (t-t_0) F(t) g(t) V(\bar{t}, t) dt}{- \int_{0}^{\infty} (t-t_0) F(t) g(t) \ln (g(t)) dt}
\raisetag{\baselineskip}
\end{equation} 
for the mean typicality assumption, and
\begin{equation}
Y(\bar{t})_{\mathrm{med}} = \frac{g(\bar{t})  - \alpha \int_{\bar{t}}^{\infty} F(t) g(t) V(\bar{t}, t) \sgn(t_0 - t)  \, dt}{\int_{0}^{\infty} F(t) g(t) \ln(g(t)) \, \sgn(t_0 - t) \, dt}
\raisetag{\baselineskip}
\end{equation} 
for the median.

In equation 10, the second (negative) term, and its derivative, approach zero as $\bar{t}$ approaches $t_0$, while the $Y(\bar{t})$ term approaches a non-zero, positive value from above, so there is guaranteed to be a window of time before $t_0$ when making the universe more hostile will increase the expected number of visible civilizations.  Figure 1 shows that this window can extend over nearly the entire history of advanced life in the universe.  For a completely general variation, $\delta E_{obs}$ could be positive or negative, depending on the shape of $F(t)$ and the time over which $\delta F(t)$ is concentrated.  

\begin{figure}[t]
	\centering
	\subfloat[]{
		\includegraphics[width=0.9\linewidth]{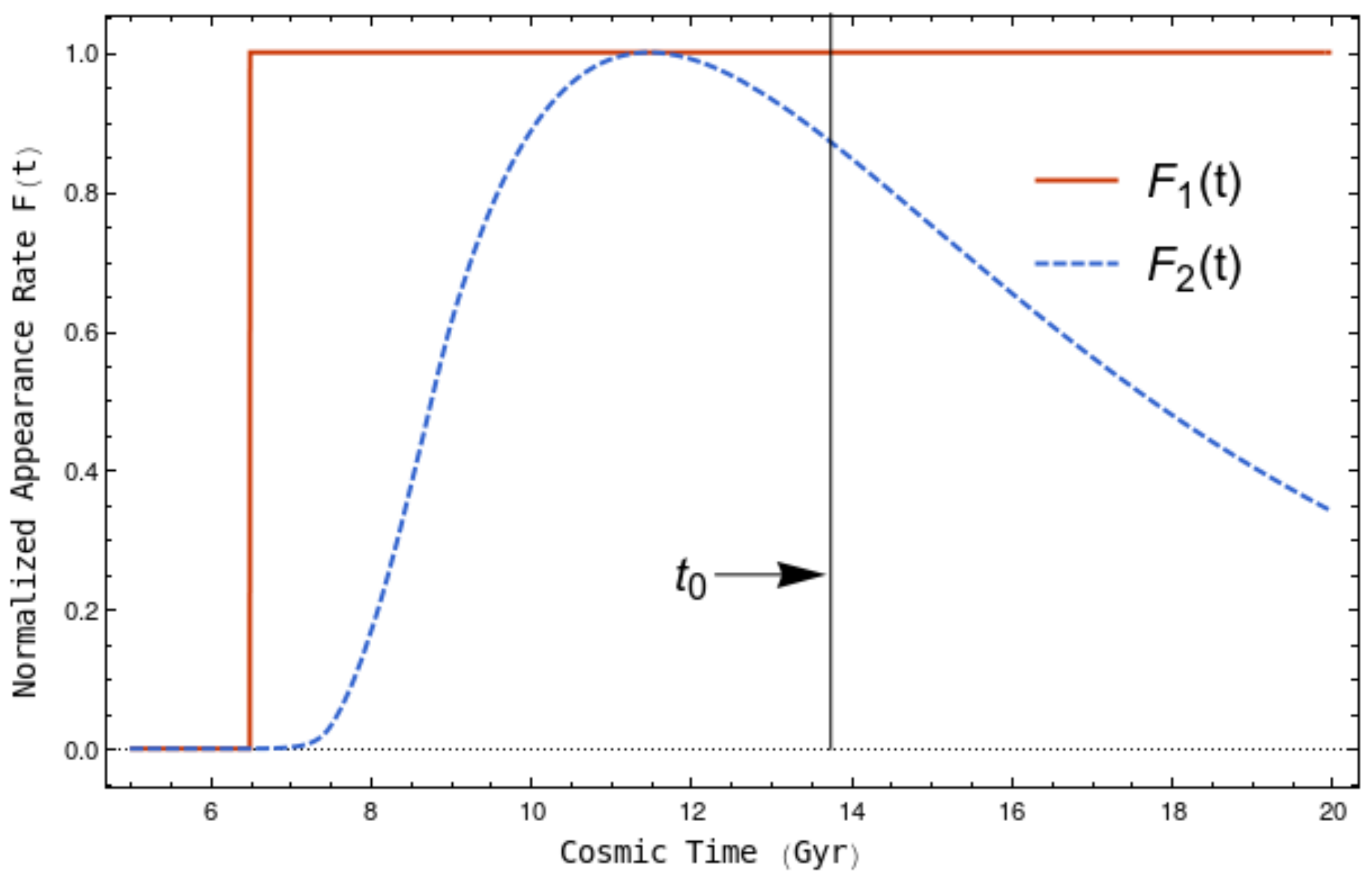}
	}
	\hspace{0mm}	
	\subfloat[]{
		\includegraphics[width=0.9\linewidth]{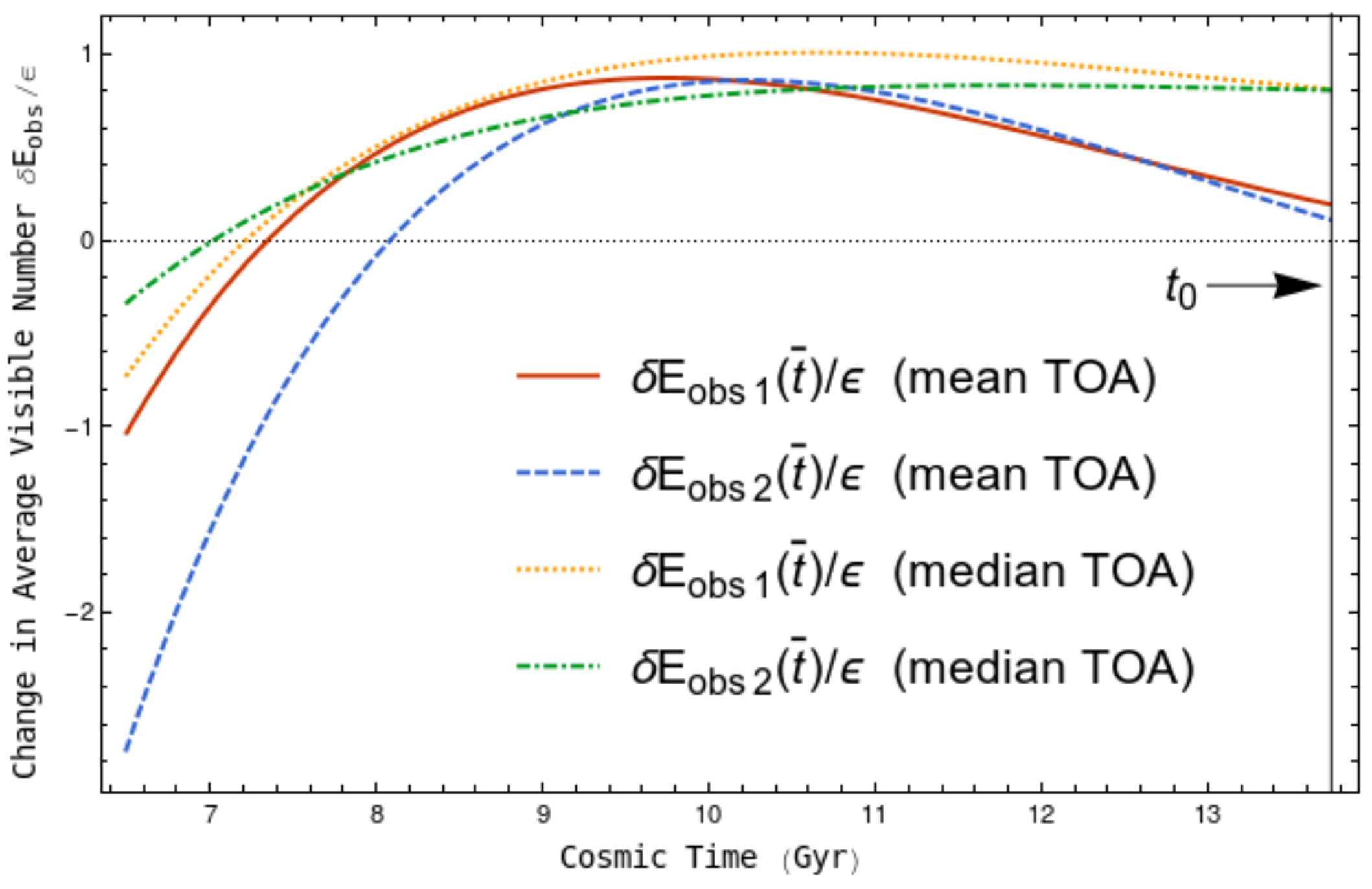}
	}
	\caption{Two appearance-rate models are given in (a), with respect to which (b) shows four responses in the expected number of visible K3+'s induced by an ``impulsive risk function" concentrated at cosmic time $\bar{t}$, i.e. $\delta F(t) = - \epsilon \, \delta(t - \bar{t})$.  Both mean and median time of arrival assumptions are shown.  Introducing danger at $\bar{t}$ increases the expected number of visible K3+'s for nearly the entire time that life could plausibly have arisen in the past.  Displacement model used is given by equation 5, with $v=.3$c.}
\end{figure}

Similar results are seen for other useful typicality assumptions, e.g. average plus $n$ standard deviations, last $q$-quantile, etc. though the full extent of the effect is unknown.  Our result is a demonstration that a counterintuitive effect is present in scenarios with strong displacement, but the boundaries are yet to be studied.

Four examples of $\delta E_{obs}(\bar{t})/\epsilon$ are depicted in figure 1b, using mean and median typicality assumptions and two different baseline functions for $F(t)$, depicted in figure 1a.  $F_1$ is a step function, switching on a $t= 6.5$ Gyr, while $F_2$ is a more realistic life appearance rate (of the type used in~\cite{olson2014,olson2017a,olson2016}).  All cases assume a $v = .3 c$ expansion/displacement model described by equation 5. One can see that for both rates, $\delta E_{obs}$ responds positively to an impulsive risk function at $\bar{t}$ over nearly the entire time that advanced life could have arisen in the past.  Only at the earliest possible times will introducing danger result in reduced estimates for the number of observable K3+'s.  This means that a \emph{realistic} risk function, spread over all earlier times, would need to have very special properties to result in lower estimates for K3+ visibility.

Previous published results used a rate similar to $F_2$ from figure 1a, and modulated it with two increasingly severe models of extinction events at earlier times, suppressing early values of $F$~\cite{olson2017a}.  In every single case out of 81 scenarios, including 3 different typicality assumptions (mean time of arrival, mean + $\sigma$, and mean + $2 \sigma$) and 9 different expansion velocities, moving from a relatively ``safer" to a ``more dangerous" early universe resulted in larger estimates for the number of visible K3+'s.

In an earlier context, where the study of high-energy extinction events was focused on the Milky Way galaxy itself, an observer-selection effect was repeatedly emphasized by Vukoti{\'c} and {\'C}irkovi{\'c}~\cite{vukotic2007,cirkovic2008a,vukotic2008}, who pointed out that newly proposed galaxy-scale extinction events should not dissuade traditional SETI searches within the Milky Way.  The idea being that if high-energy events have recently become rare enough for us to appear, others may be appearing at nearly the same time.  Indeed, they went further to point out that such extinction models actually undermine earlier anthropic arguments~\cite{carter1983} for SETI contact pessimism.  However, they took early life-hostile conditions due to gamma ray bursts to imply a ``complete lack of any extragalactic intelligent signals or phenomena"~\cite{cirkovic2008a} -- the opposite of the conclusions here.  The crucial difference in our analysis is made by including the displacement of habitable worlds.  Although our arguments are quite different in scope and structure, we find the parallels (and contrasts) interesting. 

In passing, we note that higher estimates for the number of visible K3+'s do not imply the estimates are large in an absolute sense.  It is not difficult to propose parameters that result in $E_{obs}$ that is substantially less than one.  For example, this will be the case if mature technology makes $v \gtrsim .9 c$ expansion easy enough that it becomes the dominant expansionistic behavior~\cite{olson2017a}.  The anthropic effect described here mainly shows that early hostile conditions cannot be a self-contained explanation for the Great Silence, as far as extragalactic life is concerned.

\begin{acknowledgments}
I am pleased to acknowledge Milan {\'C}irkovi{\'c} for his helpful comments and suggestions.  
\end{acknowledgments}

\bibliography{ref5}{}
\bibliographystyle{apsrev4-1}

\end{document}